\documentclass[aps,twocolumn, superscriptaddress, svgnames]{revtex4-2}
\usepackage[T1]{fontenc} 
\usepackage{tikz,tikzscale, relsize, pgfplots, xcolor, amssymb, amsmath, siunitx}
\usetikzlibrary{backgrounds, arrows, calc, decorations, positioning, decorations.pathmorphing,decorations.pathreplacing, 
	decorations.markings, fixedpointarithmetic, fit, patterns, decorations.text, backgrounds,matrix,plotmarks, spy, arrows.meta, shapes.geometric}
\usepgfplotslibrary{fillbetween}
\pgfplotsset{compat=1.17}
\usepackage{graphicx}
\usepackage{subcaption}
\usepackage{ragged2e}
\usepackage{xcolor}

\begin{document}
    
\title{Electron-beam-induced modification of gold microparticles in an SEM}

\author{Kristina Weinel}
\email{k.weinel@ifw-dresden.de}
\affiliation{Federal Institute for Material Research and Testing, Unter den Eichen 87, 12205 Berlin}
\affiliation{Leibniz Institute for Solid State and Materials Research Dresden e.V., Helmholtzstr. 20, 01069 Dresden}
\affiliation{Institute for Solid State and Materials Physics, Technische Universit\"at Dresden, Haeckelstr. 3, 01069 Dresden}
\author{Marc Benjamin Hahn}
\affiliation{University of Potsdam, Karl-Liebknecht-Str. 24-25, 14476 Potsdam}
\author{Axel Lubk}
\affiliation{Leibniz Institute for Solid State and Materials Research Dresden e.V., Helmholtzstr. 20, 01069 Dresden}
\affiliation{Institute for Solid State and Materials Physics, Technische Universit\"at Dresden, Haeckelstr. 3, 01069 Dresden}
\author{Wen Feng}
\affiliation{Leibniz Institute for Solid State and Materials Research Dresden e.V., Helmholtzstr. 20, 01069 Dresden}
\author{Ignacio Gonzalez Martinez}
\affiliation{Leibniz Institute for Solid State and Materials Research Dresden e.V., Helmholtzstr. 20, 01069 Dresden}
\author{Bernd B\"uchner}
\affiliation{Leibniz Institute for Solid State and Materials Research Dresden e.V., Helmholtzstr. 20, 01069 Dresden}
\affiliation{Institute for Solid State and Materials Physics, Technische Universit\"at Dresden, Haeckelstr. 3, 01069 Dresden}
\author{Leonardo Agudo J$\acute{\text{a}}$come}
\affiliation{Federal Institute for Material Research and Testing, Unter den Eichen 87, 12205 Berlin}

\begin{abstract}
Electron-beam-induced conversion of materials in a transmission electron microscope uses the high power density of a localized electron beam of acceleration voltages above 100 kV as an energy source to transform matter at the sub-micron scale.
Here, the e-beam-induced transformation of precursor microparticles employing a low-energy e-beam with an acceleration voltage of 30 kV in a scanning electron microscope is developed to increase the versatility and efficiency of the technique. 
Under these conditions, the technique can be classified between e-beam lithography, where the e-beam is used to mill holes in or grow some different material onto a substrate, and e-beam welding, where matter can be welded together when overcoming the melting phase.
Modifying gold microparticles on an amorphous SiO$_x$ substrate reveals the dominant role of inelastic electron-matter interaction and subsequent localized heating for the observed melting and vaporization of the precursor microparticles under the electron beam. Monte-Carlo scattering simulations and thermodynamic modeling further support the findings.
\end{abstract}

\maketitle

\section{Introduction}
The strive for downsizing tools, devices and materials while maintaining or improving performance, or generating new functionality, has been, amongst others, supported by the development of electron-beam(e-beam)-based fabrication. For instance, e-beam lithography \cite{Thompson1983, Tseng2003}, welding \cite{Elmer1990, Siddharth2020} or additive manufacturing \cite{Engstrom2014, Gibson2021}, have been utilized down to the submicron-scale for the fabrication of devices such as micro-electromechanical systems (MEMS) and nano-electro-mechanical systems (NEMS) from a very broad selection of materials \cite{Maluf2004}. However, further reducing length scales as well as the energy and material footprint of the manufacturing process remains one of the challenges.

The recent advancement of targeted microparticle modification via an e-beam in a transmission electron microscope (TEM), which utilizes beam energies between 80kV and 300kV, has added another technique to this toolbox, which is capable of structuring and synthesis at the nanoscale thanks to the high precision of electron optics in a TEM \cite{GonzalezMartinez2016,Caldwell2010, Longo2013}. A particular appeal of this technique is the possibility of studying the as-produced materials such as nanoparticles (NPs) in terms of structure and composition directly in-situ in the TEM, providing an untarnished view of the synthesis products without further sample modification or preparation. Despite the notable advances in this method, a full understanding of the e-beam-induced material modification process has not been accomplished yet as the impact of e-beam-induced heating and charging depends on a cascade of scattering as well as heat and charge diffusion processes. These complicated processes are entangled evading a precise and comprehensive description.

Consequently, exploring the synthesis at lower e-beam energies such as those used in a scanning electron microscope (SEM) would be highly beneficial to reach a better understanding and also wider applicability of the technique. In addition to the roughly one order of magnitude lower acceleration voltage, the SEM is characterized by its scanning mode, which complicates the e-beam matter interaction, due to the spatio-temporal modulation of the interaction. Notwithstanding, a couple of synthesis techniques are adopted to exploit the interaction of an electron beam with matter in the SEM. For example, e-beam lithography is used for milling a pattern into different materials on nanoscopic length scales, e.g. photo-resists. Electron-beam-induced deposition is a technique in which the electrons are used to decompose evaporated molecules whose heavy atoms deposit on a substrate. Electron beam welding uses the molten phase of matter to either separate material or weld it together. The e-beam-induced modification of microparticles, however, does operate in a somewhat different parameter space than these techniques, concerning the introduced heat, charge and induced modifications. 

In the following, we explore the use of SEM to induce a material transformation with the electron beam operating at low beam energies as an alternative to hitherto implemented TEM protocols. The working principle of electron-beam-induced modification in the SEM is exemplarily shown for gold microparticles (MPs) deposited on amorphous silicon oxide (a-SiO$_x$) substrate as a benchmark precursor system. Typical observations, including the production of gold nanoparticles, are shown, and the underlying interaction processes are addressed in a first approximation. 
Subsequently, we start with a description of the experimental methods employed, followed by detailed Monte-Carlo scattering simulations as well as thermodynamic modeling to reveal the driving mechanisms behind the observed material modifications.

\section{E-beam-induced microparticle modification}

\subsection{Scanning beam protocol}
\label{sec_Protocol in SEM}

To prepare the precursor on the substrate, the following procedure was applied to spherical gold MPs (Alpha Aesar, Thermo Fisher Scientific) whose nominal diameter $d$ ranges from $d=\SI{0.8}{\mu m}$ to $d=\SI{1.5}{\mu m}$, although particles of more than $d=\SI{2.9}{\mu m}$ were found in practice. The MPs were dry-sprinkled onto a copper grid (200 mesh) coated with a holey amorphous mixture of SiO and SiO$_2$ (SPI Supplies), which will be referred to as "a-SiO$_x$" from now on. The mean thickness of the substrate, $t=(\SI{52}{} \pm \SI{16}{})\SI{}{nm}$ was measured by SEM as well as by Electron Energy Loss Spectroscopy
in TEM. 

The scanning beam protocol (sBP) for electron-beam-induced material modification was implemented in a Quanta 3D FEG 200/600 (FEI Company) SEM. All the secondary electron (SE-)SEM images, referred to as "frames" in the article, were acquired with the built-in Everhart Thornley (ET-) Detector at room temperature (RT) and pressure of $P \approx \SI{3}{\times 10^{-4}\  Pa}$. To produce a precise beam current control and a high beam current stability, the analytical mode was employed, which limits the beam current to the range of discrete values between $I=\SI{5}{nA}$ and $I=\SI{48}{nA}$. 
These beam currents have been calibrated by using a Faraday cup. 

\begin{figure}[htbp]
    \centering    
    \includegraphics[width=8cm]{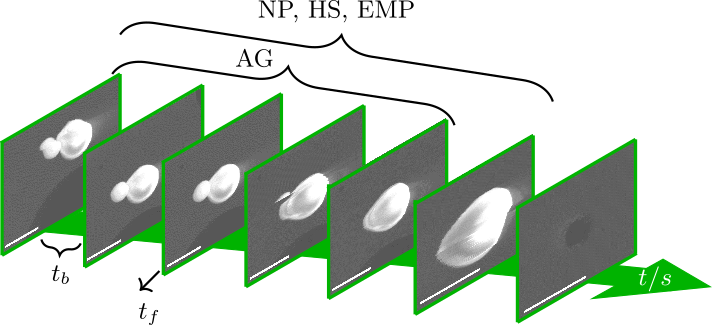}
\caption{The fabrication process is resolved in time. In this example, an e-beam with $V=\SI{30}{kV}$ and $I=\SI{11}{nA}$ was used. Aggregation (AG) occurred first after 5 frames and further irradiation of the MP with the e-beam leads to nano-particles (NP) in the vicinity of the occurred hole in the substrate (HS) and the ejection of the MP (EMP). Scale bar corresponds to $=\SI{2}{\mu m}$, $t_b$ is the time where the beam is blanked which is normally in the range of $t_b>\SI{15}{s}$, $t_f$ is the frame time which is $t_f<\SI{1}{s}$.
}
    \label{Im_Time-resolution_Process}
\end{figure}

A schematic of the sBP is presented in Fig. \ref{Im_Time-resolution_Process}. The figure shows a series of frames that are recorded at constant beam parameters, such as acceleration voltage, current or working distance, during the experiment. In this example, $V=\SI{30}{kV}$ and $I=\SI{11}{nA}$ were used. Between the frames, the beam was blanked. The scan time for each frame is $t_f<\SI{1}{s}$, whereas the time while the beam is blanked is $t_b>\SI{15}{s}$, that is at least one order of magnitude longer than the scan time. The magnification may change from frame to frame, i.e. the area of the scan changes. We note that the number of frames considerably varied from experiment to experiment, mainly due to differences in the number of frames required to observe a final MP state.

By scanning the beam multiple times over the gold MP, different types of MP modifications, referred to as "outcomes" in the following, can be observed as shown in Fig. \ref{Im_Time-resolution_Process}. Details on the outcomes will be given in section \ref{sec_summarizing the outcomes}). 
An experiment, i.e. series of frames with the same current, of a single MP started at low e-beam current. In the case, that no visible MP transformation is observed, a second series of frames with a higher e-beam current was applied to eventually induce a modification.
More than 60 gold MPs on a-SiO$_x$ substrate were irradiated, resulting in a set of outcomes hinting to at a stable and repeatable behavior of gold MPs on a-SiO$_x$ substrate under the applied e-beam conditions.
  
We finally note that, while all the mentioned e-beam parameters and material properties and geometries do influence the e-beam-induced modification, it will be shown below that they are not sufficient to describe the final state in a deterministic way. This is likely attributed to other unknown boundary conditions, such as the substrate microparticle contact area, which are not-well defined but as also important. We therefore abstain from a more exhaustive experimental study of these parameters.

\subsection{Modification results}\label{sec_summarizing the outcomes}

\begin{figure}[htbp]
    \centering    
    \includegraphics[width=8cm]{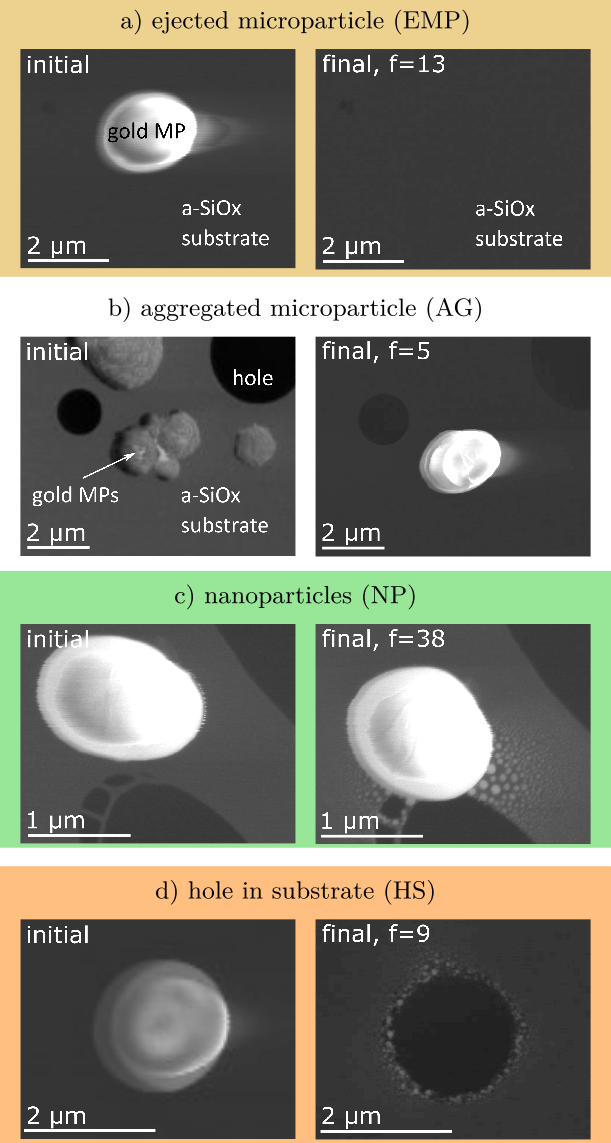}
\caption{Four not necessarily exclusive outcomes: \textbf{a)} ejection of MPs (EMP),  \textbf{b)} aggregation of MPs (AG), \textbf{c)} formation of nanoparticles (NP),  and \textbf{d)} formation of a hole in the substrate (HS).}
    \label{fig_NumberOfReaction}
\end{figure}

The observed outcomes were evaluated and categorized to distinguish the various states of MP modifications that can be produced. 
The systematic study is based on more than 60 experiments at locations with at least one initial gold MP. At some locations, more than one outcome was observed.
At least four different kinds of outcomes are identified, as shown in Fig. \ref{fig_NumberOfReaction}.

\begin{itemize}
 \item Ejected microparticle (EMP) (Fig. \ref{fig_NumberOfReaction}a)): gold MPs can be ejected.
 The push-off of the MPs from the substrate itself occurs on time scales faster than the time resolution of the experiment (order of seconds), which can be attributed to the negative charging of the MP and the substrate, repelling each other.
 \item Aggregation (AG) (Fig. \ref{fig_NumberOfReaction}b)): if more than one initial MP, which are in contact with each other, are illuminated, they tend to fuse to form a new single gold MP. This type of outcome is typically observed before other outcomes, in accordance with previous reports where Au MPs are illuminated by an external energy source \cite{Ru1997,Akman2013}.
 \item Nanoparticles (NP) (Fig. \ref{fig_NumberOfReaction}c)): NPs were found in the surrounding of the gold MP, while it simultaneously shrinks. \cite{Ru1997,Akman2013} At higher applied currents, the process starts already during the first frame. The physical reasons of this behavior of gold under an e-beam is still not conclusively clarified and will be discussed below in more detail.
 \item Hole in the substrate (HS) (Fig. \ref{fig_NumberOfReaction}d)): holes were created in the a-SiO$_x$ substrate. If holes were produced also NP were found in the surroundings, but not vice versa. This type of outcome is attributed to severe substrate heating.
\end{itemize}
This list of outcomes is perhaps not complete but accessible by using the ET-detector in the SEM.
                            
\begin{figure}
    \centering    
    \includegraphics[width=8cm]{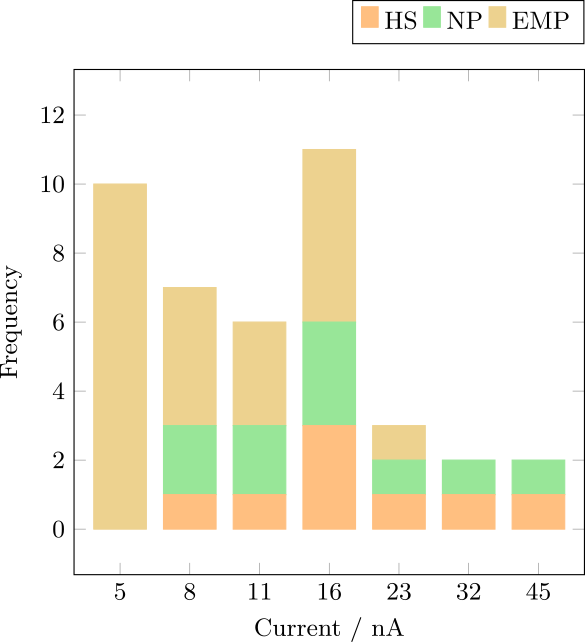}
\caption{Relative frequency of the outcomes regarding the current of the applied electron beam at constant acceleration voltage of $V=\SI{30}{kV}$.}
    \label{fig_Histogram}
\end{figure}

To investigate how the e-beam current influences the type and frequency of outcomes, the acceleration voltage and the scanning parameters are kept constant, while the applied current changes in the range between $I=\SI{5}{nA}$ and $I=\SI{45}{nA}$. The result is shown in Fig. \ref{fig_Histogram} for three acquired outcomes EMP, NP and HS. These experiments started with one initial MP.  

At the low current of $I=\SI{5}{nA}$, only EMP can be observed, which hints at charging effects that are insufficiently compensated, e.g., by electron currents through the substrate. The frequency of EMP reduces by increasing the current, and the outcomes of NP as well as HS become more probable. At a current of $I=\SI{8}{nA}$, the synthesis of gold NPs as well as HS can be observed which reach their main frequency at $I=\SI{16}{nA}$. Fewer modifications in total are observed at higher currents because the treatment of the MPs started at low currents, reaching a series of frames with higher currents only in the case of no observable outcome.

The occurrence of the aggregation of more than one initial gold MPs can be considered as a preparation step as long as it takes place as the first modification. The new aggregated single MP counts then as one initial particle for further treatment of the MP with an e-beam. Hence, Table \ref{tab_frequency of AG} shows the relative frequency of this kind of outcome regarding the applied current of the e-beam.

\begin{table}[htbp]
 \centering
 \begin{tabular}{|l|l|l|l|l|l|l|l|}
 \hline
 Current / nA&5&8&11&16&23&32&45\\
 \hline
 Relative frequency / \% of AG&0&30&40&20&10&0&0\\
 \hline
 \end{tabular}
\caption{Frequency of aggregated MPs versus current.}
\label{tab_frequency of AG}
\end{table}

At a low current of $I=\SI{5}{nA}$, no AG was observed, implying that indeed only EMP (and no visible other reactions) occur at this current. In the range between $I=\SI{8}{nA}$ and $I=\SI{23}{nA}$, AG takes place with different frequencies. 
All these observations indicate that there is a correlation between the kind of outcome and the applied current of the e-beam \cite{Egerton2004}, although several unknown and uncontrollable parameters (such as contact area of MP and sample) lead to a considerable scatter. 

\subsection{Electron-beam-induced charging} \label{sec_SE-detector}

\begin{figure}[htbp]
    \centering
    \includegraphics[width=8cm]{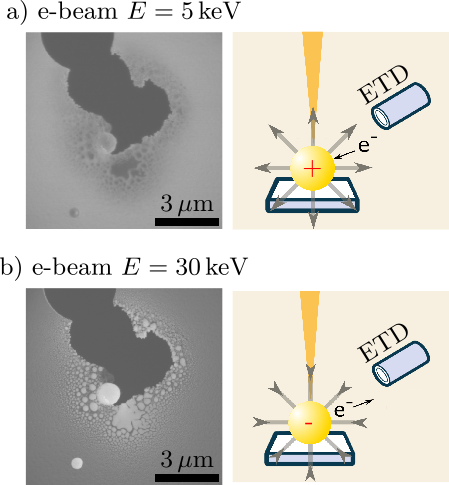}
    \caption{Comparison of ETD-SE frames of the same location, but with different e-beam energies.
    a) $E=\SI{5}{keV}$ and b) $E=\SI{30}{keV}$ show different contrasts which are related to the yield of SEs in the ETD detector at each scan point. The yield in turn depends not only on the activation of the SEs but also on the path of the electrons to the detector which differ under the influence of different additional electrostatic fields as sketched on the right-hand side.
    }
    \label{fig_SEM_ElectricField}
\end{figure}

To assess the charging of the MP by the electron beam, ETD-SE frames acquired at the same location, but with different incident energies were qualitatively compared, as exemplarily shown in Fig. \ref{fig_SEM_ElectricField}. 

\begin{figure}[htbp]
    \centering
    \includegraphics[width=8cm]{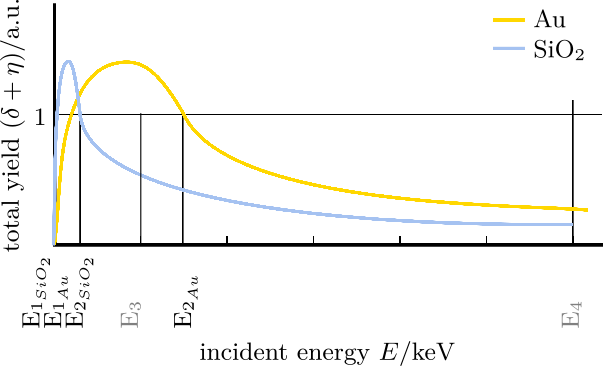}
    \caption{Sketch of the total emission yield depending on the incident electron energy for gold (yellow) and silicon dioxide (blue). 
    The values for $E_1$ and $E_2$ differ due to material properties \cite{Joy1996}.
    Regimes above 1 are positively and regimes below 1 are negatively charged. $E_{3}=\SI{5}{keV}$ corresponds to the e-beam energy used for the charging experiments.
    $E_{4}=\SI{30}{keV}$ corresponds to the applied energy in the modification experiments.
    }
    \label{fig_SEM_charging}
\end{figure}

Generally, both types of charging, positive and negative, can occur in an SEM, which mainly depends on the incident beam energy as sketched in Fig. \ref{fig_SEM_charging} and also on material properties. The number of incoming primary electrons (PE) and the number of electrons leaving the MP (SE or backscattered electrons (BSEs) for example) can compensate each other, leading to a total emission yield of 1 where no charging occurs (see $E_1$ and $E_2$ in Fig. \ref{fig_SEM_charging}). Between $E_1$ and $E_2$, positive charging takes place and for the other regimes of incident energies, negative charging occurs. \cite{Rosenkranz2011, Schmidt1994}

Typical values for $E_1$ are several $\SI{}{eV}$ depending on the material, beam current, vacuum etc.
The value of $E_{2_{Au}}\approx \SI{7.5}{keV}$ for gold was found in literature.
$E_{2_{SiO_2}}$ for silicon oxide could not be found, but with the approximation $E_2 \approx 0.12\ Z$ with $Z$ the atomic number, it could be calculated to be $E_{2_{SiO_2}} = \SI{1.2}{keV}$ with $Z=10$ the mean atomic number. In conclusion, the values for $E_{2_{Au}}$ and $E_{2_{SiO_2}}$ differ a lot and the e-beam of $E_3=\SI{5}{keV}$ is located in between pointing to a positively charged gold MP and a negatively charged substrate since the a-SiO$_x$ substrate follows this theoretical bulk behavior. \cite{Joy1996}\\

Fig. \ref{fig_SEM_ElectricField}a) represents a SE frame recorded with an incident beam energy of $E_3=\SI{5}{keV}$ and shows a relatively low contrast which can be explained by an electric field around the gold in which emitted electrons are attracted back to the MP and less SE emission is detected in the ETD. Therefore, the field lines point outwards (see the sketch on the right-hand side of Fig. \ref{fig_SEM_ElectricField}a)), which implies that the gold MP is positively charged.

Fig. \ref{fig_SEM_ElectricField}b) represents a SE frame recorded with an incident beam energy of $E_4=\SI{30}{keV}$ and shows a relatively high contrast, which can be explained by an electric field around the gold MP in which emitted electrons are repelled from the MP and higher SE yield can be detected in the ETD. Therefore, the field lines point in the direction of the MP (see the sketch on the right-hand side of Fig. \ref{fig_SEM_ElectricField}b)), which implies that the gold MP is negatively charged.

\section{Scattering and thermodynamic simulations}

Interaction between the e-beam and matter includes a multitude of physical processes which can be classified as elastic and inelastic scattering, where the cross-section of each interaction depends mainly on the e-beam energy. \cite{Reimer1985}

To address the underlying physical mechanisms driving the MP modification, two numerical simulations were carried out. At first, we used Monte-Carlo scattering simulations of electrons impinging on a spherical gold MP standing on a SiO substrate to determine the sign of the charging as well as the magnitude of the deposited energy. 
Second, the latter was used to estimate the temperature increase of the MP via thermodynamic modeling.

\subsection{Monte-Carlo scattering simulations} \label{sec_result_MonteCarlo}

For the scattering simulations, we utilized the \textit{Geant4} 10.06.p03 framework and the \textit{TopasMC} 3.7 interface, employing the \textit{Livermore} scattering models as well as the \textit{"g4em-standard\_SS"} physics list with deactivated multiple-scattering processes. \cite{agostinelligeant42003a,perltopas2012} The following additional processes were activated: Fluorescence, Auger, AugerCascade, and PIXE, thereby taking into account the large majority of inelastic scattering processes at low beam energies with the notable exception of (volume) plasmon excitation. The latter results in an underestimation of the stopping power. The energy loss due to bulk plasmons is namely low compared to other losses (notably core electron excitation). 

To evaluate the scattering results regarding the spatial distribution of electron and deposited energy, the particle, whose diameter is of $d=\SI{1.9}{\mu m}$, was cut into 1\,nm slices on all three spatial dimensions. The gold  and SiO have a density of $\rho_{Au}=\SI{19.32}{\frac{g}{cm^3}}$ and $\rho_{SiO}=\SI{2.1}{\frac{g}{cm^3}}$, respectively. A detailed discussion of the particle scattering simulation of microscopic and nanoscopic gold structures can be found in the work by Zutta and Hahn.\cite{zuttavillateradioactive2019} For scoring of the localized energy deposit, the ``parallel-world'' feature was used. 

\begin{figure}
    \includegraphics[width=8cm]{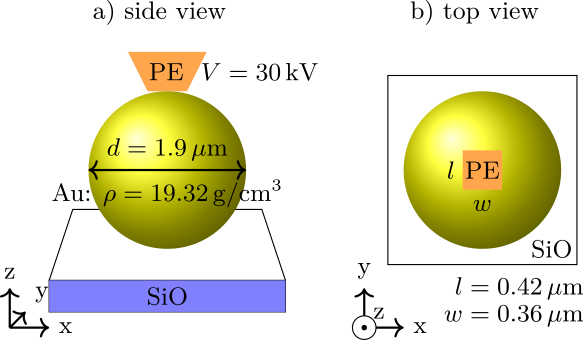}
\caption{Input geometry, material and beam parameters for the Monte-Carlo simulation.}
    \label{fig_Simulation_geometry}
\end{figure}

Fig. \ref{fig_Simulation_geometry} illustrates the geometry and the material parameters as a side view in a) and as a top view in b). The acceleration voltage of the PE is $V=\SI{30}{kV}$ and the amount of PEs is set to $1 \times 10^7 e=\SI{1.602}{pC}$ yielding sufficiently small statistical bounds to the obtained average values for energy loss, secondary electron yield, etc.

Two case studies were simulated to cover different magnification settings as used during the experiments, i.e., the relation between the scan field and the diameter of the MP:
\begin{enumerate}
    \item The scan field has a length of $l=\SI{0.42}{\mu m}$ and a width of $w=\SI{0.36}{\mu m}$, which corresponds to a scan area that is narrower than the MP, as shown by the orange region in Fig. \ref{fig_Simulation_geometry}b) in the top view as an orange box.
    \item The scan field has a length and a width of $l=w=\SI{4}{\mu m}$, which corresponds to a broad scan area compared to the MP as used for the experiment shown in Fig. \ref{Im_Time-resolution_Process} for example.
\end{enumerate}

The amount of deposited charge and energy relative to the incoming number of electrons are shown in Table \ref{tab_SimulationResult}. It is clearly shown that the amount of deposited charge $C_{\text{eff}}$ and deposited energy $E_d$ is much higher for the narrow e-beam compared to the broad e-beam case. In both cases, the deposited charge is negative, meaning that more electrons are deposited into the gold MP than emitted as secondary electrons or transmitted and backscattered electrons. This is in good agreement with the theory presented in Fig. \ref{fig_SEM_charging} and the negative charging observed experimentally (see Sec. \ref{sec_SE-detector}).

\begin{table}[htbp]
 \centering
 \begin{tabular}{|l|l|l|}
 \hline
 Property&1. narrow &2. broad \\
 & e-beam&e-beam\\
 \hline
 $C_{\text{eff}}$& 31\% & 4\%\\
 $E_{\text{d}}$& 47\%& 7\%\\
 \hline
\end{tabular}
\caption{Percentage of incoming charge, $C_{\text{eff}}$, and energy, $E_{\text{d}}$, that remains deposited in the MP for the narrow and broad e-beam cases.}
\label{tab_SimulationResult}
\end{table}
       
\begin{figure}[htbp]
    \centering
    \includegraphics[width=8cm]{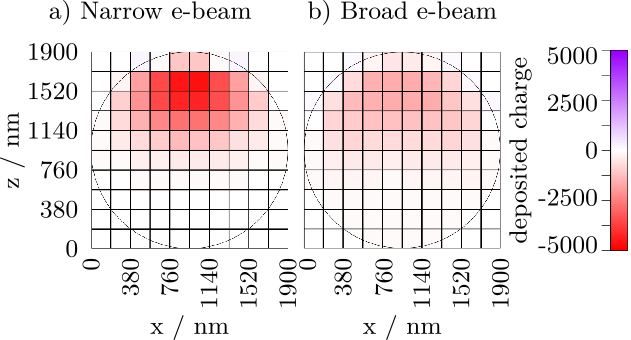}
    \caption{Deposited charge distribution in the MP by using a static simulation. a) narrow beam, as sketched in Fig. \ref{fig_Simulation_geometry}. b) broad beam.}
    \label{fig_effective Charge}
\end{figure}

The slice cut along the y-direction, shown in Fig. \ref{fig_effective Charge}, represents a two-dimensional x-z-plane through the middle of the MP, which reveals the deposited charge distribution. The e-beam impinges the MP from the top.
Fig. \ref{fig_effective Charge}a) shows the spatial resolution for a narrow e-beam. The negative (red) deposited charge indicates the penetration volume of the electrons into the MP. The lower hemisphere, where the MP contacts the substrate, shows negligible deposited charge, reflecting that most of the electrons have been stopped in the upper part. 

Fig. \ref{fig_effective Charge}b) shows the distribution of the deposited charge for a broad e-beam. The absolute amount of the deposited charge is reduced concerning the narrow e-beam because the same amount of incoming electrons is distributed over a larger area, as also shown in Table \ref{tab_SimulationResult}. Again, the negative (red) deposited charge in the MP is mostly distributed in the upper hemisphere where the e-beam enters the MP.

In total, the scattering simulations predict that $31\%$ of the incoming electrons are deposited as charge in the gold MP, imparting $47\%$ of their energy in that process. 

In the next sections, we further elaborate on the heating and charging introduced by the deposited energy and charge, respectively, considering also dissipation processes which were not taken into account in the above scattering simulations.

\subsection{Thermodynamics} \label{sec_thermodynamic}

Heat is transferred to the MP as a result of a cascade of inelastic scattering processes between PEs and matter, e.g., starting with the excitation of a core electron, which subsequently decays into phonons eventually involving intermediate excitation of electron-hole pairs, etc. In general, this is a transient process that involves a couple of sub-processes, which are partly unknown. For instance, the amount of secondary electrons ejected from the MP will change over time as the sample is charged in the electron beam (see Sec. \ref{sec_SE-detector}). To describe the heating of the MPs, we will therefore focus on the most important processes, facilitating a semi-quantitative description of the thermodynamics of the MPs under the electron beam.

The thermal energy (heat) $W$ transferred to and from the precursor MP can be decomposed into different physical processes and reads
\begin{equation}
    W=W_{\mathrm{PE}}-W_{\mathrm{SE}}-W_{\mathrm{BSE}}-W_{\mathrm{S}}-W_{\mathrm{R}}
\end{equation}
where $W_{\text{PE}}$ is the heat input through PEs; $W_{\text{SE}}$ and $W_{\text{BSE}}$, the heat dissipation through the SEs and BSEs leaving the gold MP; $W_{\text{S}}$, the heat dissipation through the substrate; and $W_{\text{R}}$, the heat dissipation through radiation (radiation losses). 

A large fraction of PEs with the kinetic energy of $E=\SI{30}{keV}$ from a narrow e-beam, that fully hits the MP, stopped within the MP (see Section \ref{sec_result_MonteCarlo}) and transferred approximately 47$\%$ of their energy (which includes already the energy removed by BSEs and SEs) mostly in the form of heat. Therefore, we can neglect the heat dissipation through the substrate because the area of the contact point of the MP with the substrate is of the order of $\mathrm{nm}^2$, which prevents an effective transfer of heat. Indeed, the contact point serves as a fuse, which eventually is responsible for burning a hole into the substrate (outcome HS) in reality. Radiation losses, however, constitute an important source of heat dissipation at elevated temperatures due to their $T^4$ dependency (Stefan–Boltzmann law), which must not be neglected.

As a consequence, we obtain a transient heating of the gold MP, which starts with the electron beam hitting the sample, where it raises the temperature, which in turn increases radiation losses until the beam leaves the sample (during the scanning process) or the sample undergoes a phase transition, e.g., melts (outcome AG) or vaporizes (outcome NP). Neglecting for a moment the possibility of phase transitions, we approximately model this process of raising the temperature by equating the change of internal energy $U$ (and hence temperature) of the MP with the net heat transferred to the MP

\begin{align}
    \frac{dU}{dt}&=\frac{dW}{dt} \nonumber \\
    & \approx \frac{I(t)\ E_{\mathrm{D}}}{q_e} - 4\pi r^2 \epsilon_\mathrm{Au} \sigma \left( T^4(t) - T_0^4(t) \right) ,
\end{align}
where $I(t)$ is the primary electron current on the MP, $E_\mathrm{D}=\frac{A}{16\mathrm{\mu m}^2}0.07\times30 \mathrm{keV}$ the energy deposited by an electron beam extended over an area $A$ (see Table \ref{tab_SimulationResult}), $q_e$ the charge, $r$ the radius of the MP, $\epsilon_\mathrm{Au}$ the emissivity of the gold MP surface, $\sigma$ the Stefan Boltzmann constant, $T$ the temperature and $T_0$ the temperature of the surrounding. The latter was introduced to take into account the heat transferred to the MP by radiation from the surrounding at room temperature, which may additionally change over time if the surroundings heat up. We note that this is only a first-order approximation, e.g., as the heating up of the surroundings (notably the substrate close to the MP) depends on the temperature $T$ of the MP in an unknown way. The increase of the internal energy of the gold MP is linearly related to its temperature increase via 

\begin{equation}
    \frac{dU}{dt}=m_\mathrm{Au}c_\mathrm{Au}\frac{dT}{dt},
\end{equation}

with $m_\mathrm{Au}$ the mass of the particle and $c_\mathrm{Au}$ its specific heat capacitance. Combining the above equations, we end up with the following non-linear first-order differential equation describing the evolution of the temperature of the gold MP

\begin{equation}
    \frac{dT}{dt}=\frac{I(t)\ E_{\mathrm{D}}}{m_\mathrm{Au}c_\mathrm{Au}q_e} - \frac{4\pi r^2 \epsilon_\mathrm{Au}\sigma}{m_\mathrm{Au}c_\mathrm{Au}} \left( T\left(t\right)^4 - \left( T_0\left(t\right)\right)^4 \right),
\end{equation} \label{equ_diff.equation}

\begin{figure}[htbp]
    \centering    
    \includegraphics[width=8cm]{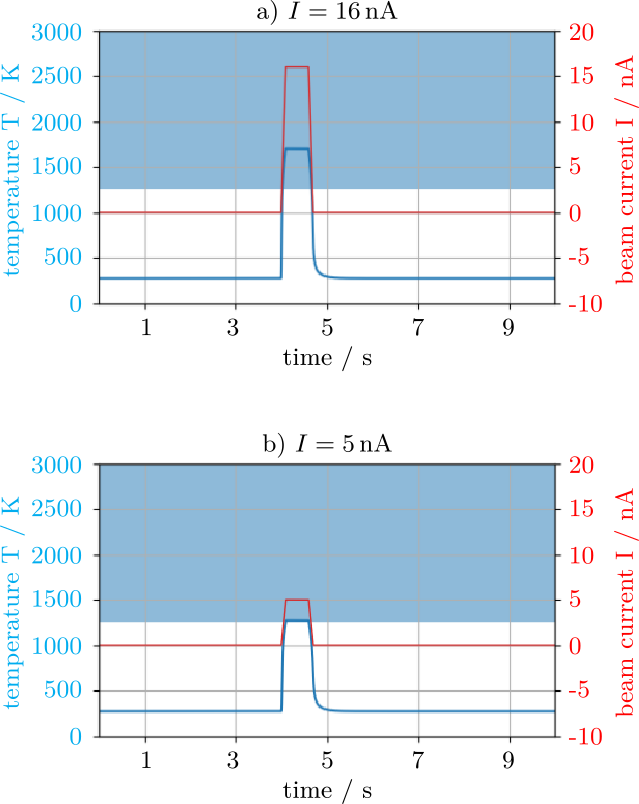}
\caption{Time evolution of temperature within one frame of $A=\SI{10}{\mu m} \times \SI{10}{\mu m}$ employing $I=\SI{16}{nA}$ (a) and $I=\SI{5}{nA}$ (b) current. The blue shaded area indicates the temperature of the gas phase according to Fig. \ref{fig_T-P-Diagram_Au}}
    \label{fig_heating}
\end{figure}

To solve the equation, we insert a time-dependent primary beam current of $I=\SI{16}{nA}$ ($I=\SI{5}{nA}$), which sweeps in $t=\SI{0.68}{s}$ over the frame of $A=10 \mathrm{}\mu m \times 10 \mathrm{}\mu m$ (see the red curve in Fig. \ref{fig_heating}a) and b), respectively). We furthermore assume that $\epsilon_\mathrm{Au}=1$ (ideal black body radiation).

Under these assumptions, the calculated temperature (blue curve in Fig. \ref{fig_heating}a) and b)) shows a strong increase of $T\approx\SI{1690}{K}$ and $T\approx\SI{1260}{K}$, respectively, while scanning a frame and a very fast dissipation of heat and hence lowering of temperature within $t=\SI{1}{s}$ due to radiation when the beam is blanked.

In a more realistic approximation, the increase of the temperature is limited by first-order phase transitions. To take that into account, a calculation of a temperature-pressure phase diagram (CALPHAD method) of gold was performed by using Thermo-Calc \cite{Andersson2002} and the SGTE Substance database \cite{Pisch2023}.

\begin{figure}[htbp]
    \centering    
    \includegraphics[width=8cm]{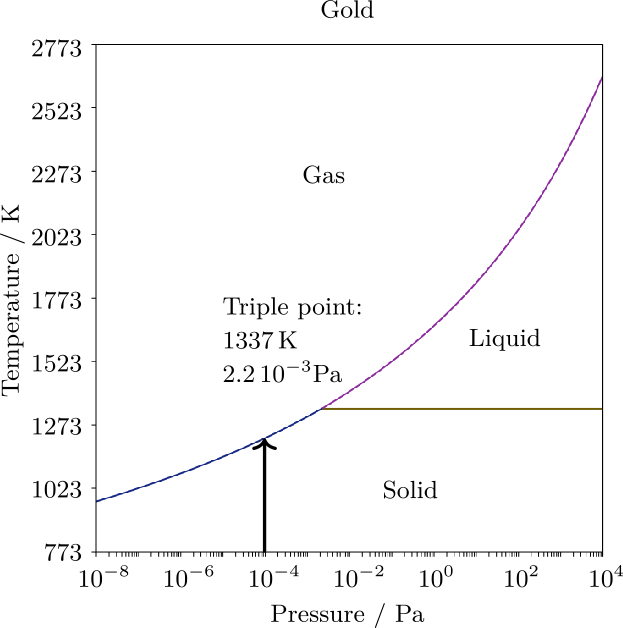}
\caption{Calculated pressure-temperature-phase diagram of gold. The black arrow denotes the pressure of $P \approx \SI{3}{\times 10^{-4}\  Pa}$ in the SEM chamber where a phase transition appears at $T = \SI{1261}{K}$.}
    \label{fig_T-P-Diagram_Au}
\end{figure}

Fig. \ref{fig_T-P-Diagram_Au} shows the calculated $P$-$T$-diagram of gold. The triple point, at which gas, liquid and solid phases co-exist with equal probability, is found for gold at $P_t=\SI{2.2}{\times 10^{-3}\  Pa}$ and $T_t=\SI{1337}{K}$. At lower pressure, only the transition between solid and gas (sublimation) takes place, whereas at higher pressure, the transition between solid and liquid (melting) occurs. The environmental working pressure within the SEM chamber is $P \approx \SI{3}{\times 10^{-4}\  Pa}$ (see the black arrow in Fig. \ref{fig_T-P-Diagram_Au}), one order of magnitude lower than the triple point. Hence, most sublimation takes place when the temperature of around $T = \SI{1261}{K}$ is reached. The gaseous gold atoms are deposited subsequently on the substrate as NPs with different distances, depending on their kinetic energy.

We therefore conclude that the temperature increase of the MP is considerably larger than the solid-gas phase transition temperature of Au in a high-vacuum environment (indicated by the shaded area in Fig. \ref{fig_heating}) in case of $I=\SI{16}{nA}$. That triggers a couple of complicated processes (vaporization of Au, Au atom convection within the MP, decomposing of the substrate), which are beyond our simple differential equation model. When inserting a current of $I=\SI{5}{nA}$ in the simulations, on the other hand, the temperature just reaches the phase transition temperature of $T=\SI{1261}{K}$. Therefore, we can distinguish two limiting regimes that are observed experimentally:

\begin{itemize}
 \item No phase transition takes place over the entire series of frames, i.e., after each temperature shock the Au MP returns to its original state. This regime is prevalent at reduced beam currents below the threshold of $I\approx \SI{5}{nA}$ where the phase transition temperature is not reached. Here, the threshold predicted by simulations agrees very well with the experimentally observed one (see Fig. \ref{fig_Histogram}).
 
 \item Thermal energy accumulation leads to an increase of the temperature beyond the phase transition temperature, triggering the evaporation of Au which redeposits on the surrounding substrate. The latter very efficiently heats the surrounding substrate, which may also accumulate over subsequent frames. 
\end{itemize}

\subsection{Charging}

To elaborate the influence of charging similarly to the heat case, we decompose the charge $Q$ transferred to and from the precursor MP into charging processes

\begin{equation}
Q =	Q_\mathrm{PE} - Q_\mathrm{S} - Q_\mathrm{SE} - Q_\mathrm{BSE},
\end{equation} \label{equ_ChargeBalance}

with $Q_\mathrm{PE}$ the incoming charge from the PEs, $Q_\mathrm{S}$ the  electrons leaving the precursor MP through the substrate, $Q_\mathrm{SE}$ the secondary electrons which have energies $< \SI{50}{eV}$ per definition when they leave the matter, and $Q_\mathrm{BSE}$ the back-scattered electrons which have energies $> \SI{50}{eV}$. \cite{Joy1996,Egerton2004,Rosenkranz2011,Zhang2022} 

As already shown in Fig. \ref{fig_SEM_ElectricField}, Fig. \ref{fig_SEM_charging} and supported by the Monte-Carlo-simulation (see Fig. \ref{fig_effective Charge}), the MP precursor is negatively charged, i.e., the sum of incoming electrons is larger than the sum of outgoing ones. 

In contrast to thermodynamic modeling, we abstain from further modeling of the transient behavior of charging because of a number of unknowns, introducing considerable errors in such a simulation. Notably, we do not know the resistivity of the MP substrate interface nor that of the substrate itself, preventing the modeling of e-beam-induced current (EBIC). Moreover, taking into account the transient charging and its impact on SE emission and SE recapture is very complicated. 

\section{Discussion and Outlook} \label{sec_discussion}

We can conclude that Monte-Carlo scattering simulations (Sec. \ref{sec_result_MonteCarlo}) in combination with simple thermodynamic modeling (Sec. \ref{sec_thermodynamic}) offers a consistent explanation for the observed MP transformations, in particular the formation of the NPs (see Sec. \ref{sec_summarizing the outcomes}). Consequently, we identify inelastic electron beam - MP interaction and the heat deposited in this process as the driving physical mechanism for the observed behavior. Above a critical current of $I=\SI{5}{nA}$, as shown experimentally in Fig. \ref{fig_Histogram}, sublimation occurs (see Fig. \ref{fig_T-P-Diagram_Au}), and the gold atoms eventually redeposit on the substrate. Alternatively, if more than one MP are illuminated with the e-beam, the induced energy leads to a high mobility of the gold atoms which in turn can lead to aggregation of the MPs (see Sec. \ref{sec_summarizing the outcomes}).
Moreover, the electron flow and thermal transport through the tiny contact area between MP and substrate can lead to holes in the substrate (see Fig. \ref{fig_NumberOfReaction}d)).
The influence of SE ejection could not be clarified, however, the good agreement between the thermodynamic calculation and the experiment suggested that SE generation exercises a rather small influence.

To obtain further insight into the physical driving mechanisms of e-beam-induced modification of MPs in the SEM,  we suggest a parameter study of the acceleration voltage of the PE as well as the MP size, thereby modifying the deposited energy, the SE generation and their balance with radiation loss and heat capacity. A further reduction of the complexity of the synthesis process, e.g., by using the e-beam in an SEM without the scanning mode, would also be helpful.

Extending this scanning beam protocol in the SEM to other precursor-substrate combinations may reveal also different types of outcomes, which in turn can widen the selection of materials utilized for the fabrication of devices at the submicron-scale, such as MEMS and NEMS. The number of suitable precursor-substrate combinations can be narrowed down by the prediction of their behavior at elevated temperature at low pressure, which was done using the thermodynamic modeling (see Sec.\ref{sec_thermodynamic}) in combination with the CALPHAD analysis (see Fig. \ref{fig_T-P-Diagram_Au}).

Possible applications for different outcomes could be the following.
Assemblies of NPs or nanodots are promising candidates for investigating the behavior of plasmons \cite{Schultz2024}, photons \cite{Schwartz2007}, phonons \cite{Ni2021} as well as acoustic surface waves \cite{Faez2009} under the influence of disordering. 
The aggregation of MPs allows the tailored growth of MPs of specific sizes, which is useful for the targeted production of MPs. Combining this capability with micro-manipulating techniques in an SEM, for example, the synthesized MP can be transferred easily to another location.
Furthermore, gold MPs of tailored sizes illuminated by an e-beam can be used as precise microscopic heating elements with tunable temperatures depending on the applied e-beam current.

\section{Conclusion}

In summary, we successfully transferred the concept of electron-beam-induced modification of precursor MPs in TEM, where a high-energy e-beam is used to modify matter, to the SEM using an e-beam of lower energy. This was exemplarily shown for gold MPs deposited on a-SiO$_x$ substrate that aggregate and/or transform into assemblies of NPs inter alia. 

We could furthermore identify the heating of the MPs stemming from the deposited energy of the e-beam via inelastic scattering as the main physical mechanism for the observed MP transformation. As a consequence of the lowered primary beam energy, the electrons are largely stopped within the MP, therefore increasing the deposited energy in comparison to high-energy electron beams utilized in a TEM for MPs of radii below several micrometers. This renders e-beam-induced modification of MPs in an SEM more efficient concerning its TEM counterpart, and hence a viable route to produce novel materials in this thermodynamically driven process.

\section*{Acknowledgements}

Dr. Nicholas Grundy, from TCSAB, is gratefully acknowledged for discussions and thermodynamic calculations. Kristina Weinel and Wen Feng acknowledge the crucial support of the funding provided by the IFW-BAM tandem program.
Axel Lubk acknowledges funding from the European Union's Horizon Europe framework program for research and innovation under grant agreement n. 101094299 (IMPRESS project).

\section*{References}

\linespread{1}
\bibliographystyle{apsrev4-2}
\bibliography{Literature_06}

\end{document}